# Fast inverse lithography based on a model-driven block stacking convolutional neural network

Ruixiang Chen, Yang Zhao, Haoqin Li, and Rui Chen[*]

*School of physics & State Key Laboratory of Optoelectronic materials and Technologies, Sun Yat-sen University, Guangzhou, 510275, China*

**Abstract:** In the realm of lithography, Optical Proximity Correction (OPC) is a crucial resolution enhancement technique that optimizes the transmission function of photomasks on a pixel-based to effectively counter Optical Proximity Effects (OPE). However, conventional pixel-based OPC methods often generate patterns that pose manufacturing challenges, thereby leading to the increased cost in practical scenarios. This paper presents a novel inverse lithographic approach to OPC, employing a model-driven, block stacking deep learning framework that expedites the generation of masks conducive to manufacturing. This method is founded on vector lithography modelling and streamlines the training process by eliminating the requirement for extensive labeled datasets. Furthermore, diversity of mask patterns is enhanced by employing a wave function collapse algorithm, which facilitates the random generation of a multitude of target patterns, therefore significantly expanding the range of mask paradigm. Numerical experiments have substantiated the efficacy of the proposed end-to-end approach, highlighting its superior capability to manage mask complexity within the context of advanced OPC lithography. This advancement is anticipated to enhance the feasibility and economic viability of OPC technology within actual manufacturing environments.

## 1. Introduction

Lithography is a crucial process in semiconductor manufacturing, responsible for transferring intricate patterns from a photomask onto a substrate, typically a silicon wafer, enabling the creation of the microscopic circuits that make up integrated circuits (ICs) and other microelectronic devices. This process involves coating the wafer with a light-sensitive material called a photoresist[1], exposing it to light through the photomask, and then developing the photoresist to reveal the desired pattern. With the ongoing demands for lower power consumption and higher integration density in ICs, the critical dimension (CD) continues to shrink. This reduction exacerbates the impact of optical diffraction and interference on lithography process, leading to a phenomenon known as optical proximity effect (OPE)[2]. OPE cause distortions in the printed pattern on the wafer, affecting the accuracy and quality of the final product[3].

To mitigate the effects of OPE, inverse lithography technology (ILT) has emerged as a vital resolution enhancement technique in optical lithography, particularly for semiconductor manufacturing at technology nodes beyond 45 nm[4]. Unlike traditional experience-based OPC methods, ILT optimizes the mask pattern as a digitized image on a pixel grid, achieving pixel-level optimization freedom[5–7]. This approach can automatically generate Sub-Resolution Assist Features (SRAFs)[8], which broadens the process window of the lithography system, ensuring improved pattern fidelity and yield.

Historically, gradient-based optimization algorithms have been the primary tools employed to address the ILT problems[9]. Ma et al. have proposed ILT optimization schemes for pixelated masks on a vector imaging model[10] and source-mask optimizations[11], as well as enhancing system robustness against variations caused by defocus, dose variation, and aberrations[12]. However, the inherent flexibility of pixel-based OPC may result in the creation of curved or intricate features that are challenging or even impossible to manufacture[13]. As a result, conventional fracturing methods require a large number of Mask Fracturing Shots to approximate the curved edges, which ultimately leads to poor manufacturability at high volumes due to the unmanageable mask writing times and accuracy degradation from jagged edge formation[14]. The Mask Fracturing Shot Count refers to the number of rectangles required to accurately reproduce the mask shape[15]. To reduce mask complexity, a TV regularization term method was proposed to reduce isolated points and burrs on the mask[16]. Furthermore, a curve optimization method based on level set implicit representation was adopted[17]. Shen et al. have introduced the level-set based mask optimization that offers enhanced the manufacturability of the mask[18].

In addition to adding regularization to the loss function, mask complexity can also be reduced by imposing structural constraints on the mask. Ma et al. have proposed a block-based OPC optimization framework that enhances the resolution of optical lithography systems, while addressing the three major manufacturing constraints and avoiding small unmanufacturable mask features induced by singular pixels[19]. Nevertheless, this block-based OPC optimization method still requires the repeated calculations of loss function gradients in each iteration, imposing a substantial computational burden. To keep pace with the increasing density of current semiconductor circuits, there is an ongoing demand for improved solver that can reduce the computational complexity of the optimization process while enhancing their applicability.

Recently, deep learning approaches have demonstrated their efficacy in addressing inverse problems across various a spectrum of applications, including imaging[20,21], optical metrology[22] and optical microscopy[23,24]. This proficiency has been also underscored by their demonstrated competence in enhancing the efficiency and precision of mask optimization, as evidenced by numerous studies [25–27]. These models, particularly deep neural networks, excels at modelling complex imaging systems and photoresist effects, which is crucial for accurate decision-making or prediction[28]. This capability has spurred the initiation of numerous deep learning projects within the field of lithography, including but not limited to Generative Adversarial Networks[29], Fourier neural operator[30], deep level-set[31], graph convolution network[32], U-net with treating ILT-guided pretraining as the major training process[13]. Given that the performance of trained models is inherently dependent on the availability of training data, and considering the significant computational resources required to acquire real datasets, Ma et al. have introduced a model-driven convolutional neural network[33,34]. This innovation approach aims to streamline the process by eliminating the time-consuming task of label generation, therefore facilitating a label-free training scheme[35].

Inspired by the aforementioned work, we developed the idea of fast inverse lithography approach based on a model-driven block stacking convolutional neural network (BSCNN). This approach enables the model to predict manufacturability-friendly OPC masks. The proposed BSCNN-ILT framework combines block stacking transmission, vector imaging models and model-driven ILT techniques. The block stacking transmission of BSCNN can inherently control the generating OPC mask manufacturability. By employing the model-driven ILT training method, the network can be trained without annotated data, and the enhanced physical model gives the neural network greater robustness. As for the training patterns, we utilize the wave function collapse algorithm to randomly generate a sufficient amount of target pattern data that conforms to the characteristics of the circuit to enhance the generalization ability of the model. The proposed BSCNN-ILT method greatly reduces mask complexity and improves the computational performance, demonstrating its ability to provide a concise and efficient solution for advanced lithography systems.

The remainder of this paper is organized as follows: Section 2 presents the vector lithography system and evaluation methods; Section 3 describes the model-driven BSCNN-ILT method for OPC; Section 4 shows the numerical results demonstrating the efficacy of the proposed method, and conclusions are drawn in Section 5.

## 2. Preliminaries

### 2.1. The vector lithography model

The lithography model includes the imaging model and the photoresist threshold model. In practical applications, the imaging model widely uses the Hopkins diffraction model[36] for partially coherent imaging systems to describe the imaging process from the mask to the wafer surface. The photoresist threshold model typically uses a Constant Threshold Resist (CTR) model to describe the development process of the photoresist under exposure conditions. Given a binary mask $M(x,y)$ where each pixel value is either 1 or 0, representing the transmittance of that region. The mask is processed through the imaging model and the photoresist model to obtain the contour printed on the wafer.

In the imaging model, let $\hat{M}(f,g)$ denote the Fourier transform of the mask $M(x,y)$. According to the Hopkins diffraction model, the spatial image of a partially coherent system in the frequency domain can be expressed as:

$$I(x,y) = \iint \iint TCC(f_1,g_1;f_2,g_2)\hat{M}(f_1,g_1)\hat{M}^*(f_2,g_2)e^{[-2\pi j(f_1-f_2)x+(g_1-g_2)y]}df_1 df_2 dg_1 dg_2 \quad (1)$$

where $(f_1,g_1)$ and $(f_2,g_2)$ describe the coordinates in the object frequency domain, and $(x,y)$ describe the coordinates in the spatial image plane. The TCC (Transmission Cross Coefficient) is the frequency domain cross-transfer function that describes the relationship between the illumination source and the point spread function of the partially coherent lithography system[37]. The asterisk $*$ denotes the conjugate operator. The TCC function is given by:

$$TCC(f_1,f_2,g_1,g_2) = \iint \tilde{\gamma}(f,g)\tilde{h}(f+f_1,g+g_1)\tilde{h}^*(f+f_2,g+g_2)\,df dg \quad (2)$$

Here, $(f,g)$ describe the coordinates in the source frequency domain. $\tilde{\gamma}(f,g)$ represents the effective source of the system, and $\tilde{h}$ denotes the optical transfer function of imaging system, which is the Fourier transform of the point spread function $h_k$.

A partially coherent lithography system can be decomposed into a sum of coherent systems. In this paper, we use the Singular Value Decomposition (SVD) method to decompose the TCC function of the partially coherent optical system, yielding the singular values $\alpha_k$ and the optical kernels $h_k$[38].

$$\alpha_k, \mathcal{F}\{h_k\} = SVD(TCC) \quad (3)$$

By squaring and then weighted summing the convolution results of the mask and the optical kernels, the aerial image intensity can be approximated as follows:

$$I = \sum_{k=1}^{N} \alpha_k \, |M(x,y) \otimes h_k(x,y)|^2 \quad (4)$$

where $h_k$ is the $k$-th optical kernel, $\alpha_k$ is the corresponding weight, and $\otimes$ denotes the convolution operator. This partially coherent lithography system is approximated by the superposition of $N$ coherent systems. In this paper, for instance, $N = 30$ means intensity in Eq. (4) equals the sum of the top 30 coherent systems with the largest weights obtained by SVD decomposition in Eq. (3).

In the context of photoresist threshold model, the hard threshold method is widely employed. According to this model, the photoresist will either be removed or retained in a specific area depending on whether the aerial image intensity $I$ exceeds the photoresist threshold $t_r$. Given that the hard threshold model is non-differentiable, a sigmoid function is often used as an approximation to the hard threshold function, particularly when computing the gradient of the loss function in mask optimization processes. Consequently, the wafer image $Z$ on the wafer can be mathematically expressed as[39]:

$$Z = \text{sigmoid}(I, a_r, t_r) = \frac{1}{1+\exp[-a_r(I-t_r)]} \tag{5}$$

We try to figure out the best photoresist threshold $t_r$ and the steepness index $a_r$ in order to obtain the least distortion wafer pattern, which is actually equivalent to controlling the intensity of the illumination. In our implementation, the photoresist threshold $t_r$ is set to 0.24 and $a_r$ is set to 100 as a constant coefficient.

### 2.2. Methods for evaluation

The quality of the mask can be measured using the pattern error (PE). Given a target pattern $\tilde{Z}$, a mask $M$, and the corresponding wafer image $Z$, the PE is mathematically defined as:

$$\text{PE} = ||Z - \tilde{Z}||_2^2 \tag{6}$$

Reducing PE helps in improving the accuracy of the wafer image on the wafer, ensuring that the final wafer image $Z$ closely matches the target pattern $\tilde{Z}$.

In addition, the OPC masks need to be fractured as a combination of rectangular variable shaped-beam (VSB) shot for mask writing[40]. In this paper, we will use the shot count of conventional fracturing to measure mask complexity and manufacturability a popular measure in optical lithography. Given the mask $M$, the VSB Shot Count refers to the number of rectangles required to accurately reproduce the mask shape:

$$\text{VSB shot count} = \text{the number of rectangles}. \tag{7}$$

A lower VSB shot count signifies that the mask can be decomposed into a reduced number of rectangles, which simplifies the manufacturing process[13].

### 3. Fast ILT based on the model-driven BSCNN

The schematic diagram of the proposed method is illustrated in Fig. 1(a), which comprises three key components: the BSCNN encoder, a model-driven decoder, and ILT training. The BSCNN encoder integrates block stacking transmission and a U-Net++ neural network. The block stacking transmission ensures that the OPC mask is composed of rectangular blocks, thereby reducing its complexity.

For a given target pattern $\tilde{Z}$, the BSCNN encoder processes $\tilde{Z}$ to generate a predicted OPC mask, $M^{pred}$. This predicted mask is then passed through the decoder for forward physical simulation, as described in the previous section, producing the corresponding wafer image $Z$. If the resulting wafer image $Z$ closely matches the target pattern $\tilde{Z}$, the predicted mask $M^{pred}$ is the desired mask pattern.

Notably, training this encoder does not require data annotation[33]. To enhance the model's generalization, a wave function collapse algorithm is employed to generate a diverse set of target patterns rich in Manhattan features, which serve as encoder inputs for training the BSCNN model. Once the training process is completed, the trained BSCNN network minimizes the error between target pattern and wafer image. At this stage, the decoder can be decoupled, allowing the BSCNN to operate independently, efficiently predicting the block stacking OPC mask ($M^{pred}$) for any given target pattern ($\tilde{Z}$). In the following, we will delve into the detail the BSCNN encoder's architecture, the preparation of target patterns and the training process of the BSCNN-ILT framework.

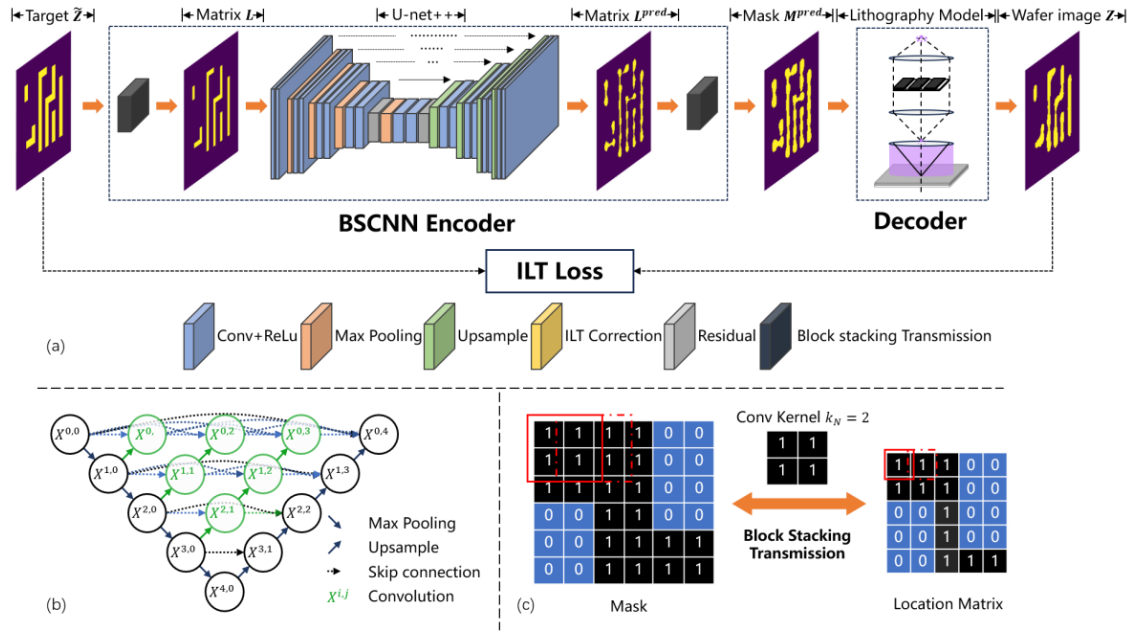

Figure 1 Overview of the BSCNN-ILT framework. (a) Model flowchart (b) U-net++ network structure (c) block stacking transmission between mask and location matrix. Here, $k_N$ is set to 2.

### 3.1. The structure of the BSCNN encoder

This section outlines the construction of the BSCNN encoder, which comprises two core components: block stacking transmission and a U-net++ neural network. The block stacking transmission ensures that

OPC masks are stacked by basic rectangular blocks, simplifying the mask structure and reducing its complexity in OPC mask.

**U-net++ Backbone**

As shown in the Fig. 1(b), the backbone of the BSCNN encoder is U-net++[41], an enhanced version of U-net. U-net++ introduces dense skip connections, enabling high-level feature maps to not only receive information from low-level feature maps but also to comprehensively utilize multi-layer feature information, improving optimization accuracy by leveraging these features during decoding. This helps better integrate features at different scales.

The U-net++ architecture employed here consists of five double convolution modules and max-pooling layers, which gradually reducing the size of feature maps while increasing the number of channels. The feature map size is then restored using up-sampling and deconvolution modules. To address potential gradient explosion issues, a residual layer is incorporated at the bottleneck, comprising two residual blocks. Each block contains two 3×3 convolution layers and batch normalization layers[42].

**Block Stacking Transmission**

To simplify the mask and ensure its manufacturability, the BSCNN encoder adopts a block stacking approach, constructing masks by stacking manufacturable rectangular blocks. As depicted in Fig. 1(c), the block stacking transmission maps the target pattern $\tilde{Z}$ to location matrix $L$ by convolving $\tilde{Z}$ with a rectangular convolution kernel $K$. The resulting location matrix $L$ represents the superposition of overlapped blocks. Rather than using a hard threshold function, the sigmoid function is applied as an approximation due to its gradient properties. The position matrix $L$ is obtained as follows:

$$L = \text{sigmoid}(\tilde{Z} \otimes K - k_N^2 + 0.5) \tag{8}$$

where the target pattern $\tilde{Z} \in \mathbb{R}^{N_m \times N_m}$ is a binary matrix. The pixel values of $\tilde{Z}$ is 0 or 1, which indicate areas without and with etching, respectively. The convolution kernel $K \in \mathbb{R}^{k_N \times k_N}$ defines the basic rectangular block, with kernel size $k_N = 2$ in Fig.1(c). The convolution process uses a stride $S = 1$, and no padding. The threshold $-k_N^2 + 0.5$ filters out the blocks that do not fully overlap with the kernel, generating the position matrix $L \in \mathbb{R}^{(N_m - k_N) \times (N_m - k_N)}$.

**Integration with U-net++**

The position matrix $L$ is then fed into the U-net++ neural network, which predicts the position matrix $L^{pred} \in \mathbb{R}^{(N_m - k_N) \times (N_m - k_N)}$ as follows:

$$L^{pred} = G(L; W_g) \tag{9}$$

where $G(\cdot)$ represents the U-net++ network and $W_g$ denotes its parameters. Subsequently, the block stacking transmission maps $L^{pred}$ the predicted mask $M^{pred}$ through deconvolution as shown in Fig.1(c), $M^{pred} \in \mathbb{R}^{N_m \times N_m}$. The goal of deconvolution is to filter out the maximum numbers, which represent the position of blocks, ensuring the predicted mask $M^{pred}$ is constructed by stacking basic rectangular blocks. The predicted mask $M^{pred}$ can be expressed as:

$$M^{pred} = \text{sigmoid}(L^{pred} \otimes K - 0.5) \tag{10}$$

The pixel values of mask $M^{pred}$ represent the transmission coefficients of the mask, where 0 and 1 respectively indicate the opaque and clear areas. The deconvolution kernel size is the same as $k_N$, with stride $S = 1$, and padding $P = 2$. This process is equivalent to placing a rectangular block, which can overlap with others, at each position where the value is 1.

**Advantages of the Encoder Design**

Unlike traditional pixel-based approach that directly generate OPC masks at the pixel level, the BSCNN encoder integrates block stacking transmission as its head and tail, with a U-net++ neural network as the core. This structure effectively captures the polygonal features of target patterns $\tilde{Z}$ and ensures the optimization of the basic unit size, reducing jagged edges and isolated points. Such design lowers VSB shot count and enhances the practicality of OPC masks. While the block stacking method introducing manufacturing constraints and a potential loss in image fidelity with increasing $k_N$, it does not significantly impact optimization accuracy since the blocks can be overlapped. It can still improve image fidelity, achieve fine patterns and address manufacturability concerns effectively.

### 3.2. Wave function collapse for data preparation

To ensure the generalization of the proposed BSCNN-ILT model, it is essential to construct a diverse and extensive dataset of target patterns. This is achieved using the wave function collapse algorithm, a procedural technique for generating complex and non-repetitive patterns or structures based on small blocks and a defined set of rules[43], as depicted in Fig. 2. In our case, the wave collapse algorithm is employed to fill a unknow pattern of $8 \times 8$ undetermined regions (CD = 45nm) with available blocks. To initialize the target pattern, the boundary parts are pre-filled with the opaque block in order to prevent the patterns from overflowing beyond the edges, which means only 7×7 regions are undetermined. The available blocks are illustrated in Fig. 2(a), and these blocks can be rotate to form 13 different patterns to make up the target pattern.

**Algorithm Rules:**
1. **Preparation:** Each undetermined region is assigned an identical wave function, which presets the possible blocks and their corresponding probabilities.
2. **Adjacency Rules:** Adjacency constraints permit blocks of the same color can be connected, while yellow transparent blocks are prohibited from being adjacent to purple opaque blocks. If a neighboring region has already collapsed, the probabilities in uncategorized regions that violates adjacency rules are set to zero.

**Algorithm Iteration:**

1. **Observation:** Identify the region with the smallest entropy (randomly selecting one if multiple regions have the same entropy). Then collapse this region based on the probabilities and fill the region with it the chosen block.
2. **Propagation:** Update the probabilities and entropy of the uncategorized regions based on the adjacency rules and the current state of the collapsed regions.

Observation is a process of collapsing the wave function of one region to a single block, chosen randomly according to the probabilities. Propagation is a process of updating the wave functions of the neighboring regions according to the rules, eliminating the probabilities of blocks that are incompatible with the observed piece. Meanwhile, the entropy or uncertainty of the neighboring regions is reduced.

The algorithm continues the upon steps until all regions have been observed or until a contradiction occurs, indicating that there is no valid piece for a particular region. Some examples of dense target patterns are shown in Fig. 2(b). By adjusting the preset generation probabilities of blocks during preparation stage, we generated 100 dense target patterns and 100 sparse patterns. The sparse features of corresponding CD are generated to ensure the model's optimization for patterns with different sparsity.

To enhance the model's generalization capabilities for patterns with varying linewidths, we augment the training paradigm through data augmentation (DA). This includes the introduction of 60 typical patterns categorized by sizes of 75 nm, 60 nm, and 52.5 nm, with 20 patterns assigned to each specific size, as illustrated in Figure 2(c). The data augmentation process involves modifying the regions in the wave function collapse algorithm to sizes of 5×5, 6×6, and 7×7, thereby improving the model's ability to simultaneously optimize patterns with multiple critical dimensions (CDs).

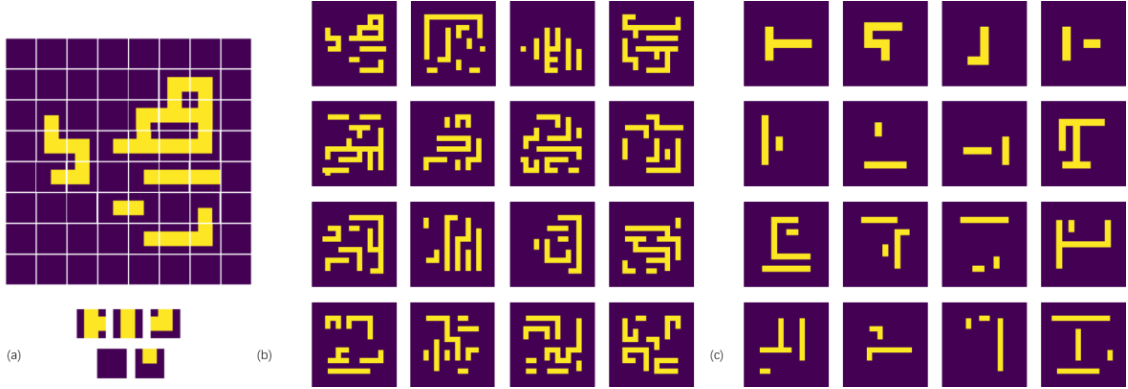

Figure 2 (a) Available pieces and a synthesized pattern example. (b) Paradigm of dense target patterns synthesized by wave function collapse. The collapse region is 8×8 and the CD is 45nm. (c) Paradigm through data augmentation (DA) with varying CDs and sparsity. From the first to the fourth rows, the collapse region is 5×5, 6×6, 7×7 and 8×8, and the CD is 75nm, 60nm, 52.5nm, 45nm respectively.

### 3.3. Training process

To obtain an optimized BSCNN encoder, the back propagation algorithm is utilized to minimize the loss function and refine the U-net++ weights $W_g$ through gradient descent:

$$\widehat{W}_g = \arg\min_{W_g}[Loss(M^{pred}, L^{pred}; W_g)] \tag{11}$$

The loss function comprises two components, defined as:

$$\text{Loss}(M^{pred}, L^{pred}) = R_{ILT}(M^{pred}) + \gamma_D R_D(L^{pred}) \tag{12}$$

The first term $R_{ILT}(M^{pred})$ is ILT correction metric, measuring the closeness of the target pattern $\tilde{Z}$ and the wafer image $Z$. The second term is the discretization penalty term, $R_D(L^{pred})$, of position matrix, which ensures that the binarity of the predicted matrix. $\gamma_D$ is the weight of $R_D(L^{pred})$. The metric $R_{ILT}(M^{pred})$ and $R_D(L^{pred})$ are expressed as:

$$R_{ILT}(M^{pred}) = \left\| Z(M^{pred}) - \tilde{Z} \right\|^2 \tag{13}$$

$$R_D(L^{pred}) = 4L^{pred^T}(1 - L^{pred}) \tag{14}$$

In this study, the network parameters are optimized based on Eq. (11) using the back propagation algorithm. During the training phase, it is necessary to compute the gradient $\frac{\partial Loss}{\partial W_g}$ for updating the mask with respect to all network weights $W_g$. The gradient of the lithography loss $R_{ILT}$ is derived using the chain rule:

$$\frac{\partial R_{ILT}}{\partial W_g} = \frac{\partial R_{ILT}}{\partial M^{pred}} \frac{\partial M^{pred}}{\partial L} \frac{\partial L}{\partial G(L; W_g)} \frac{\partial G(L; W_g)}{\partial W_g} \tag{15}$$

Where $\frac{\partial R_{ILT}}{\partial M^{pred}} \frac{\partial M^{pred}}{\partial L}$ in Eq. (15) can be derived as:

$$\frac{\partial R_{ILT}}{\partial M^{pred}} \frac{\partial M^{pred}}{\partial L} = -a_r K \otimes [\frac{\partial R_{ILt}}{\partial M^{pred}} \odot \left(\text{sigmoid}(L^{pred} \otimes K - 0.5)\right) \\ \odot \left(1 - \text{sigmoid}(L^{pred} \otimes K - 0.5)\right)] \tag{16}$$

$$\frac{\partial R_{ILt}}{\partial M^{pred}} = 2a_r\{H^T \otimes [(Z - \tilde{Z}) \odot Z \odot (1 - Z) \odot (\text{sigmoid}(L^{pred} \otimes K - 0.5) \otimes H^*) + \\ (H^T)^* \otimes [(Z - \tilde{Z}) \odot Z \odot (1 - Z) \odot ([\text{sigmoid}(L^{pred} \otimes K - 0.5)] \otimes H)]\} \tag{17}$$

$\odot$ is the element-wise multiplication, and $\otimes$ is the convolution operation.

Similarity, the gradient of discretization penalty term $R_D$ is derived using the chain rule:

$$\frac{\partial R_D}{\partial W_g} = \frac{\partial R_D}{\partial L} \frac{\partial L}{\partial G(L; W_g)} \frac{\partial G(L; W_g)}{\partial W_g} \tag{18}$$

$$\frac{\partial R_D}{\partial L} = 4 - 8L^{pred} \tag{19}$$

The ILT loss and gradient functions are built upon Eq. (4), (5), and (8), whose computational workflows closely resemble the forward and backward propagation processes in neural networks. Within this framework, the Loss function represents the forward propagation loss, while the gradient, expressed as $\frac{\partial Loss}{\partial M^{pred}} \frac{\partial M^{pred}}{\partial L}$, serves as the key component for backward propagation. The network weights $W_g$ are updated via the chain rule: $\frac{\partial Loss}{\partial M^{pred}} \frac{\partial M^{pred}}{\partial L} \frac{\partial L}{\partial G(L; W_g)} \frac{\partial G(L; W_g)}{\partial W_g}$. Furthermore, the CUDA implementation of the lithography forward model seamlessly integrates the entire forward and backward propagation process into a unified, CUDA-compatible deep learning framework. By utilizing PyTorch's automatic differentiation library (*torch.autograd*), both the lithography model and block stacking transmission can automatically track functions and tensors, facilitating efficient gradient computation throughout the optimization process.

## 4. Simulations and analysis

This section presents the simulation results, demonstrating the superiority and effectiveness of the proposed BSCNN-ILT method. The analysis focuses on the network's generalization capability for layouts generated using the wave function collapse algorithm and its performance on publicly available test sets. Further, a comparative evaluation with pixel-based methods and an assessment of the impact of wave function-based data augmentation are conducted.

### 4.1. Parameter setting

The target patterns used for training and test are randomly generated using wave function collapse algorithm, as described in Sec.3.2. Each pattern is represented at an image resolution of $98 \times 98$ pixels ($0.735 \times 0.735$ μm$^2$). The training dataset consists of 260 target patterns, while the test dataset includes approximately 130 target patterns. For the training of the neural network, we employ a batch size of 64 and conducts a total of 1,000 epochs. The Adam solver[44] is used with a step size of $1 \times 10^{-5}$. For the regularization parameter $\gamma_D$ in Eq. (12), values of to 0.1, 0.15, and 0.2 are assigned for $k_N$ of 2, 3, and 4, respectively. These configurations demonstrated improved performance in predicting OPC masks with PE and VSB shot count efficiency, leading to their adoption as the default settings for this study. The simulations are conducted on a personal computer equipped with an Intel Core i5-13600K CPU @ 3.50 GHz, 32GB RAM, and an NVIDIA RTX 3060Ti graphics card.

The lithography simulations employ a partially coherent lithography system with an illumination wavelength of $\lambda = 193$ nm. The illumination source is configured as annular aperture, with inner and outer partial coherence factors of $\sigma_{in} = 0.7, \sigma_{out} = 0.975$, respectively. The projection system was equipped with NA of 1.35. The side length of the mask pattern is $N_M = 98$ pixels, where the pixel size on mask is 7.5 nm × 7.5 nm.

### 4.2. Simulation results based on simple layout patterns

To demonstrate the superiority of BSCNN-ILT, a large amount of simple target patterns is first used for numerical experiments. For comparative analysis, the pixel-based methods with regularizations are also provided. Four distinct simple layout patterns are depicted in Fig. 3 and Fig. 4. For each pattern in the figure, the first row presents the origin mask patterns, followed by the OPC mask patterns obtained through the pixel-based method with PV-Band regularization (PVB)[45], the pixel-based method with total variation regularization (TV), and BSCNN-ILT method ($k_N$ =2,3,4). The subsequent two rows exhibit their corresponding wafer images, and the discrepancies between wafer images and target patterns, respectively. It should be note that the pixel-based method utilizes the same network structure is just the proposed BSCNN-ILT method with $k_N = 1$. For the pixel-based method with PVB (Pixel-PVB), a regularization weight of 0.0002 is applied. The pixel-based method with TV regularization (Pixel-TV) employs a TV weight of 0.00001. Both methods undergo training for 800 iterations until convergence is achieved, using a learning rate of 0.0001. The BSCNN-ILT method ($k_N$= 2, 3, 4) shares the same network structure and is trained for 1000 iterations with a learning rate of 0.0001. In these numerical experiments, a trial-and-error methodology is employed to optimize the hyperparameters for each model.

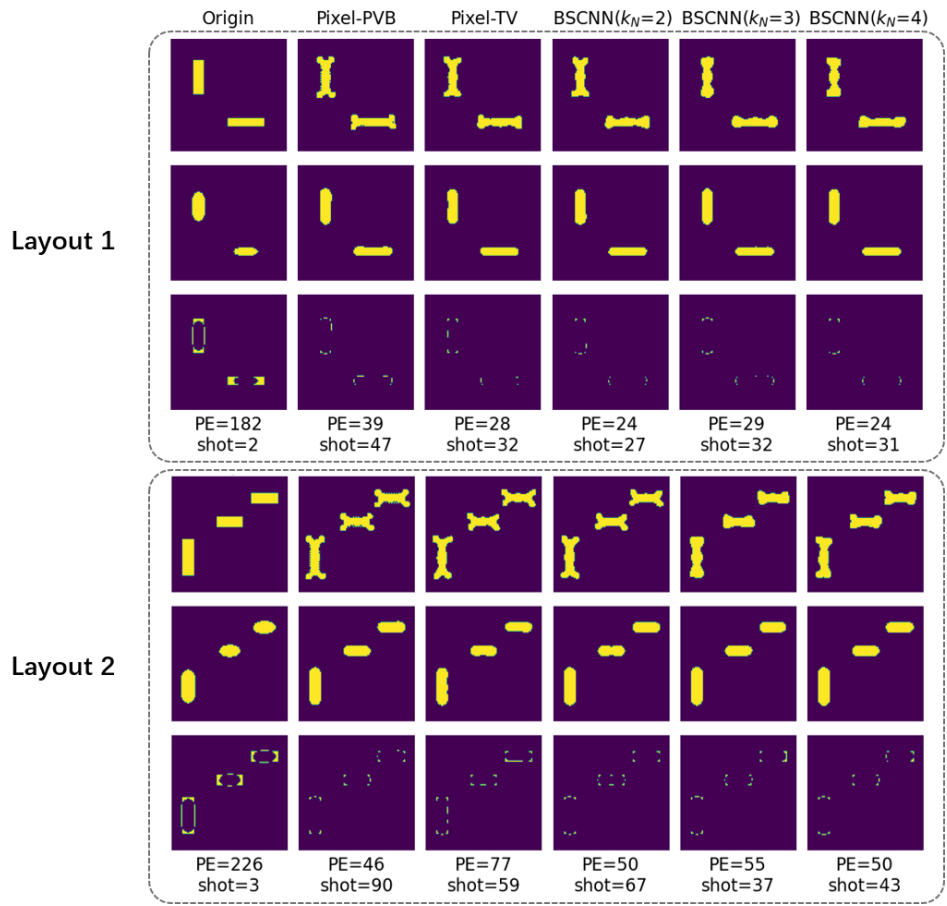

Figure 3 Simple layout pattern 1 and 2. The pattern is composed of simple horizontal rectangles, vertical rectangles. The first row presents the mask patterns, followed by the OPC mask patterns obtained through the pixel-based with PVB, pixel-based with TV, and BSCNN-ILT method ($k_N$ =2,3,4). The second and third rows display their corresponding wafer images, and the error between wafer images and target patterns, respectively.

As depicted in Fig.3 and Fig.4, the pronounced image distortions observed in the wafer images (The second row of each layout pattern) within the lithography system can be effectively compensated for all the pixel-based ILT methods. However, the complexity of the resulting mask pattern varies considerably. For the first layout pattern (Layout Pattern 1, as shown in Fig.3), the proposed BSCNN-ILT method yields mask patterns that have the analogous PE to those produced by other pixel-based ILT methods with a minor reduction VSB shot count. This outcome is anticipated, as the layout pattern is relatively simple with slight OPE to compensate, comprising two well-separated rectangles, one vertical and the other one horizontal.

Compared to the first layout pattern, the second layout pattern introduces an additional rectangle between the existing two rectangles. It has been observed that the proposed BSCNN-ILT method ($k_N = 3$) achieves a competitive PE value while generating a simpler mask pattern compared to other methods. This is evidenced by the reduction of VSB shot count to below 40 when PE remains around 50. As a result, the proposed BSCNN-ILT method yields results that remain essentially comparable PE and achieve less VSB shots to those obtained from traditional pixel-based ILT method with regularizations.

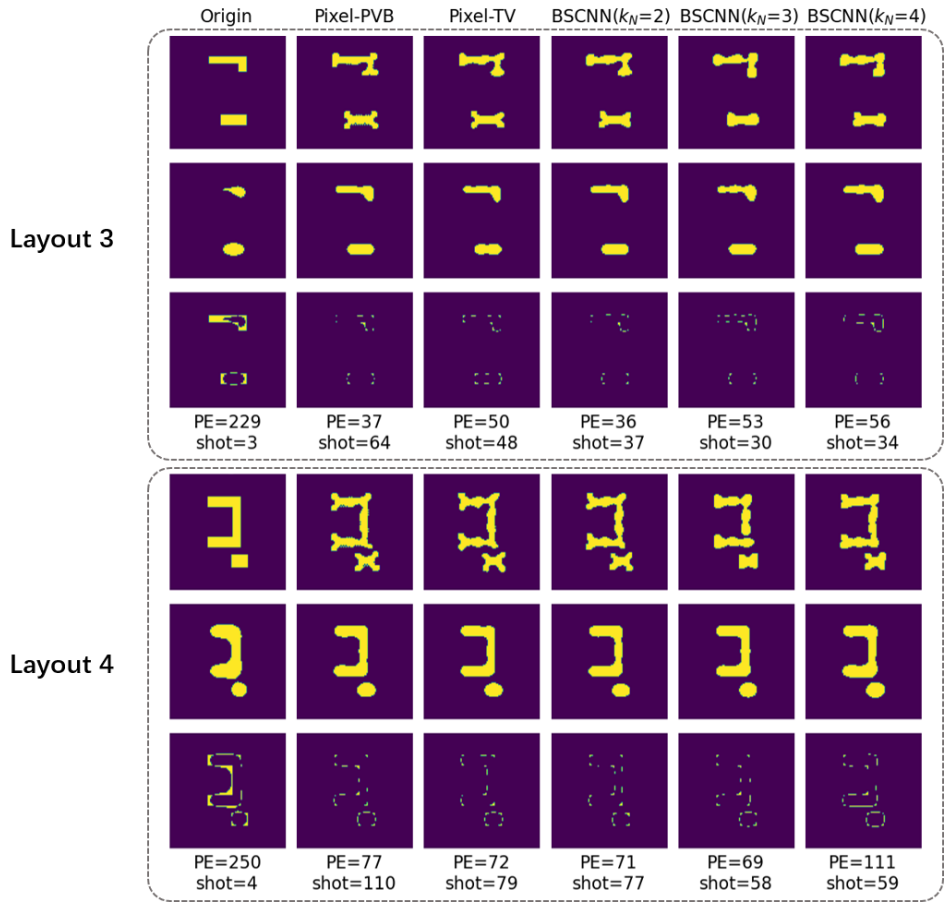

Figure 4 Simple layout pattern 3 and 4. The pattern is composed of simple horizontal rectangles, vertical rectangles and corners. The order of the images is the same as that in Fig.3.

Furthermore, the inclusion of complex corner structures is incorporated in layout pattern 3 and 4, as shown in Fig.4, highlights the adaptability of the proposed BSCNN-ILT method. As expected, the mask patterns generated using this method exhibit an obvious lower VSB shot count compared to other pixel-based ILT methods (such as Pixel-PVB and Pixel-TV), thereby indicating a simple manufacturing process. However, it is important to note that the increasing the block stacking kernel size entails a tradeoff, as the simplification of the mask pattern correspondingly leads to an increase in the PE value, especially for a target pattern with severe OPE. Thus, a balance must be maintained between the complexity of the mask and the accuracy of the generated mask patterns, highlighting the need for careful consideration in the selection of blocking stacking kernel size to optimize lithographic performance. To better explore the impact on balancing PE and VSB shot count, we will investigate it on complex layout patterns.

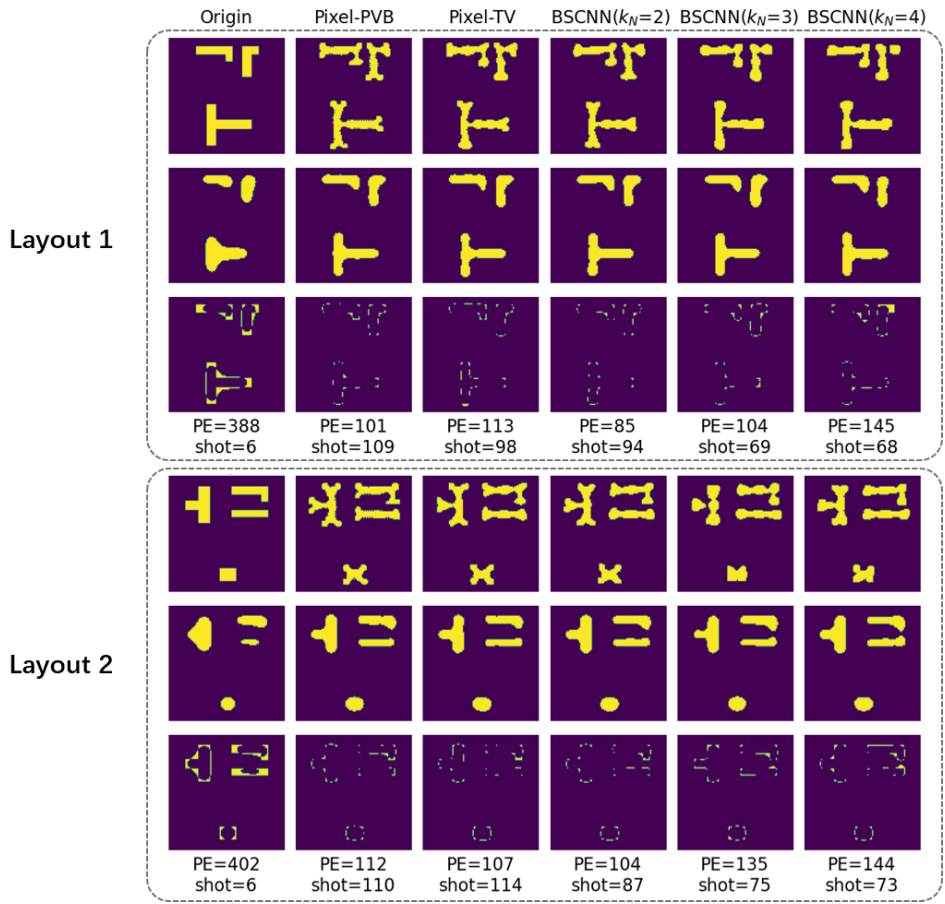

Figure 5 Complex layout pattern 1 and 2. These two layout patterns is from open-source dataset, which are composed of multiple horizontal rectangles vertical rectangles and T-shapes. The first row presents the mask patterns, followed by the OPC mask patterns obtained through the Pixel-PVB, Pixel-TV, and BSCNN-ILT method ($k_N$=2,3,4). The second and third rows display their corresponding wafer images, and the error between wafer images and target patterns, respectively.

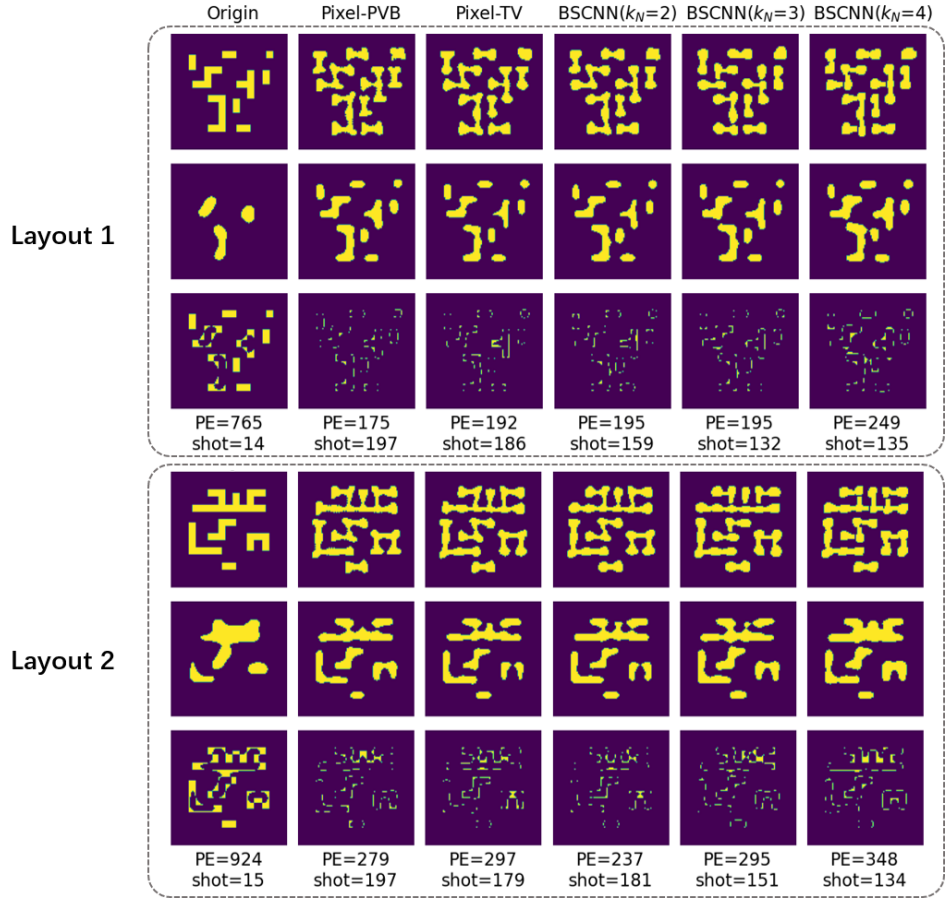

Figure 6 Complex layout pattern generating by wave function collapse. The pattern is composed of multiple horizontal rectangles, vertical rectangles, and various types of corners. The order of the images is the same as in the above figure.

### 4.3. Simulation results based on complex layout patterns

To demonstrate the generalization capability of BSCNN-ILT, the proposed model was tested and analyzed using complex layout patterns. These patterns were derived from two sources: an open-source dataset[46] and a generated dataset using wave function collapse algorithm. Both datasets encompass a wide range of typical featured found in integrated circuit layouts. Four examples of the patterns from these datasets are illustrated in Fig.5 and Fig.6. Specifically, Fig.5 depicts two distinct layout patterns from the open-source dataset, while Fig. 6 depicts presents two complex patterns of generated through wave function collapse algorithm. For each pattern, the top row presents the origin mask patterns, followed by the OPC mask patterns obtained through Pixel-PVB, Pixel-TV, and BSCNN-ILT methods with varying kernel sizes ($k_N$=2,3,4). The second row presents the corresponding wafer images, while the third row displays discrepancies between wafer images and target patterns.

Compared to simple patterns in Fig.3 and Fig.4, the complex layout patterns shown in Fig. 5 consist of six rectangles with corners and T-shapes, exhibiting significant OPE. As seen from these results, the BSCNN-ILT model ($k_N = 2$) achieves the smallest PE while maintaining a VSB shot count comparable to Pixel-PVB and Pixel-TV methods. For $k_N = 3$ and 4, the VSB shot count is further reduced, albeit at the expense of a slight increase in PE. This trade-off is manageable and highlights the BSCNN-ILT model's robustness in handling complexity. Similar phenomena can be found in the generated layout patterns depicted in Fig. 6, although they feature even more complex and dense structures, comprising multiple horizontal and vertical rectangles as well as various corners configurations.

To further evaluate the effectiveness of the proposed BSCNN-ILT method, Figure 7 illustrates the statistical analysis of PE and VSB shot count across a large dataset of simple and complex layout patterns. For simple patterns (denoted in black), the BSCNN-ILT method ($k_N = 3$) achieves the lowest VSB shot count, followed by BSCNN-ILT ($k_N = 4$) and BSCNN-ILT ($k_N = 2$). Specifically, $k_N = 3$ configuration reduces VSB shot counts to 66% of the Pixel-TV method while incurring only a 5% increase in PE. Thus the proposed BSCNN-ILT method ($k_N = 3$) offers an improvement in manufacturability without compromising precision. In the contrast, while $k_N = 4$ configuration reduces the VSB shot count to only 75% of the Pixel-TV method, it increases PE to about 1.1 times. This indicate that larger kernel size does not necessarily lead to better results.

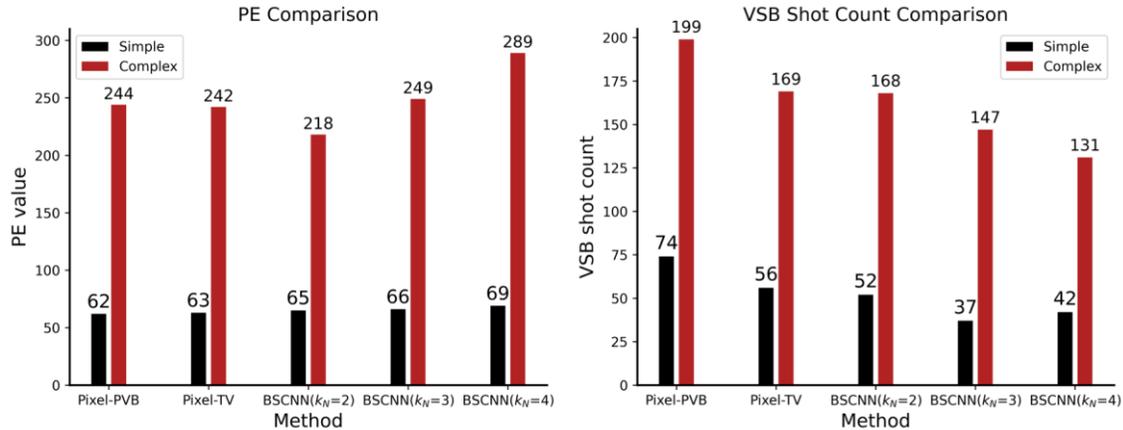

Figure 7 Statistical results of VSB shot count and PE under different methods. From left to right are the Pixel-PVB, the Pixel-TV, BSCNN-ILT method ($k_N$= 2, 3, 4), respectively. The black legend represents the statistical results for simple target patterns, while the red legend represents the statistical results for complex patterns

For complex patterns (denoted in red), the BSCNN-ILT method ($k_N = 4$) achieves the lowest VSB shot count, followed by $k_N = 3$, $k_N = 2$, the pixel-TV method, and the Pixel-PVB method, while the BSCNN-ILT method ($k_N = 2$) achieves the lowest PE, followed by the pixel-TV method, the Pixel-PVB method and the BSCNN-ILT method ($k_N = 3,4$). Specifically, the BSCNN-ILT method ($k_N = 2$) achieves a similar VSB shot count to the Pixel-TV method while reducing PE by 10%. When we increase the kernel size of block stacking to $k_N = 3$, the VSB shot count is reduced to about 87% of the pixel-TV method while only a 3% increase in PE. However, $k_N = 4$ achieves the lowest VSB shot count (about 78% of the Pixel-TV method) but at the cost of a 19% increase in PE. These finding indicate that the proposed BSCNN-ILT method is effective to reduce the complexity of the masks but we need to choose a suitable kernel size.

Therefore, in practical applications, a moderate block size ($k_N = 2$ or 3), are recommended to avoid excessive sacrifice in PE. By dealing with block-stacking rather than individual pixels, the BSCNN-ILT method inherently reduces isolated pixel points, resulting in a significant decrease in VSB shot count and improved mask manufacturability. This block-level optimization leverages the larger spatial contribution of image blocks compared to individual pixels.

**4.4. Evaluation of data augmentation**

In integrated circuit layouts, it is common for both minimum CD patterns and those exceeding the minimum CD to coexist. To this end, it is imperative to incorporate patterns larger than minimum CD as target patterns for the encoder during training within the data augmentation framework. This approach significantly enhances the model's predictive capabilities of the patterns associated with varying nodes.

To assess the impact of data augmentation on model generalization, we compared the performance with and without data augmentation (DA). The mask generation-based data augmentation strategy used to create an augmented dataset is detailed in Section 3.2. This augmentation dataset includes 100 basic dense patterns and 100 extra sparse patterns, all with a CD of 45 nm, along with another 60 typical patterns measuring with sizes of 75 nm, 60 nm, and 52.5 nm, with 20 patterns allocated to each of these size categories. In contrast, the basic dataset consists solely the 100 dense patterns with a CD of 45 nm, without any patterns larger than CD.

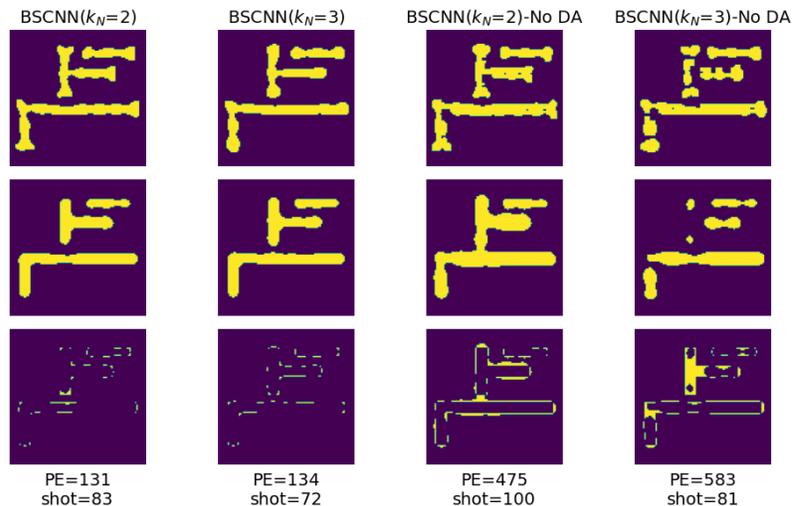

Figure 8 OPC mask and results comparison trained based on datal augmentation dataset and basic dataset. From left to right are BSCNN-ILT with data augmentation ($k_N = 2, 3$), BSCNN-ILT without data augmentation ($k_N = 2, 3$), respectively.

To validate this approach, Figure 8 illustrates the neural network predictions for two models trained on different datasets: (1) the basic dataset without data augmentation, and (2) the augmented dataset generated using the method outlined in Section 3.2. Both models employed the BSCNN-ILT configuration described in Section 4.1. The results reveal that the BSCNN-ILT model, when trained with data augmentation, demonstrates improved generalization ability across various CD features. These findings suggests that the augmented backbone model offers greater optimization effectiveness and practical applicability compared to its conventional counterpart.

## 5. Conclusion

We propose a model-driven BSCNN-ILT approach that effectively optimizes the manufacturing of OPC masks by integrating block stacking transmission, vector imaging models and model-driven ILT techniques. This innovative approach leverages a model-driven ILT training methodology, enabling the neural network to be trained without the need of annotated data. By incorporating an advanced physical model, our method enhances adaptability and robustness through the training process. To support the training procedure, we employ the wave function collapse algorithm to randomly generate a diverse and sufficiently complex dataset of target pattern that reflect the characteristics of circuit layouts. Numerical experiments have confirmed the effectiveness of the proposed end-to-end approach.

Unlike traditional methods that employ regularization terms to balance optimization weight between ILT precision and pattern complexity, the BSCNN-ILT method eliminates the need for careful weight selection, providing a clear framework to balance optimization complexity and accuracy. This complexity-reduction process can also be extended to other neural network models. It also highlights the need for careful consideration in the selection of blocking stacking kernel size to optimize lithographic performance. In conclusion, the BSCNN-ILT method proves effective to produce precise and manufacturable mask designs with an optimal kernel size. These advantages make it a concise and effective solution for addressing the challenges of advanced lithography systems, offering a robust pathway toward enhanced manufacturability and performance.

**Acknowledgments**

The author acknowledges the support of the National Key Research and Development Program of China (2022YFA1404304); Basic and Applied Basic Research Foundation of Guangdong Province, China (2023B1515040023, 2020A1515010626); National Natural Science Foundation of China (61905291).